\documentclass[oldversion]{aa}
\usepackage{txfonts}
\usepackage{natbib}
\usepackage{graphicx}

\begin{document}
\title{Radio variability properties for radio sources}
\author{J. H. Fan\inst{1,2} \and Y. Liu\inst{1} \and Y. H. Yuan\inst{1}
\and T.X. Hua\inst{1} \and H. G. Wang\inst{1} \and Y. X. Wang\inst{3}
\and J. H. Yang\inst{4} \and A. C. Gupta\inst{5} \and J. Li\inst{1}
\and J.L. Zhou\inst{1} \and S.X. Xu\inst{1} \and J.L. Chen\inst{1}}

\offprints{J. H. Fan, \email{fjh@gzhu.edu.cn}}

\institute{ Center for Astrophysics, Guangzhou University, Guangzhou
510006, China \and Physics Institute, Hunan Normal University,
Changsha, China \and College of Science and Trade, Guangzhou
University, Guangzhou 511442, China \and Department of Physics and
Electronics Science, Hunan University of Arts and Science, Changde
415000, China \and Tata Institute of Fundamental Research, Homi
Bhabha Road, Colaba, Mumbai - 400005, India }

\date{Received 6 June 2006 / Accepted ??}

\abstract{In this paper, we used the database of the university of
Michigan Radio Astronomy Observatory (UMRAO) at three (4.8 GHz, 8.0
GHZ, and 14.5 GHz) radio frequency to analyze the radio light curves
by  the power spectral analysis method in search of possible
periodicity.  The analysis results showed that the radio sources
display astrophysically  meaningful periodicity ranging from 2.2 to
20.8 years in their light curves at the three frequencies. We also
calculated the variability parameters and investigated the
correlations between the variability parameter and the flux density.
For the variability parameters, we found that the parameters at
higher frequency are higher than those in the lower frequency. In
addition, the variability parameters of BL Lacertae objects are
larger than those of flat-spectrum radio quasars. suggesting  that
they are more variable than flat spectrum radio quasars.}


\maketitle

\section{Introduction}

The nature of the central engine of blazars and other classes of
active galactic nuclei (AGNs) is still an open problem. Blazars'
light curves were generated by using the data of their monitoring
program, which have  yielded very valuable information about the
mechanisms operating in these sources and important implications for
quasar modeling \citep{fan98}. In the past two decades, optical
monitoring program of blazars and other classes of AGNs have been
conducted extensively by many groups around the globe, and blazars
were reported to display flux variability on diverse time scales
ranging from a few minutes to hours, to days, to months, and to even
more than 10 years \citep{fan05b}. The variability time scale on
years gives the long-term variation information in the source and an
important tool predicting other outburst times.

Radio monitoring programs were carried out at Bologna at 408 MHz
\citep{bondi96}, Michigan University at 4.8 GHz, 8 GHz, 14.5 GHz,
and Mets$\ddot{a}$hovi observatory at 22 GHz, 37 GHz, 87 GHz,
 ESO site on Cerro La Silla, Chile, at 90 GHz and 230
GHz \citep{torn96}. Using the data of these monitoring, many groups
have investigated the variability properties and found that blazars
show interesting results.  Based on this radio data base,
\citet{kraus99} report that blazars' emission  is strongly beamed,
\citet{laht99} estimated the radio Doppler factors for a sample of
radio sources, \citet{ciar04} investigated the possible periodicity
for several selected objects, and \citet{aller03} investigated other
variability properties.

In this paper, we make use of the UMRAO data base to investigate the
possible periodicity, the variability parameter, and the correlation
between the source brightness and the variability parameter. Section
2 presents the details about periodicity analysis method, Sect. 3
variability parameter, Sect. 4, the results of the present work, and
in sect. 5 our discussions and conclusions are given.

\section{Power spectral (Fourier) periodicity analysis method and results}

There are many methods of time series data analysis. The fact that
the astronomical observations are generally not evenly sampled will
put some constraints on the analysis methods. In this paper, we used
the power spectral analysis to search  periodicity
  in the radio light curves of
radio sources   because it is a powerful and familiar method for
detecting a periodic signal, and it gives some quantitative criteria
for the detection of a periodic signal.

Many attempts at power spectral analysis have been made for  the
case that the data are unevenly spaced in time. the {\em modified
periodogram}  was widely used by astronomers \citep{scar82,horne86},
it is based on a least-square regression onto the two trial
functions, $\sin(\omega t)$ and $\cos(\omega t)$. A superior
technigue is the {\em date-compensated discrete Fourier
transform},or DCDFT \citep{ferr81,fost95}, a least-square regression
on $\sin(\omega t)$, $\cos(\omega t)$ and constant.  The DCDFT is a
more powerful method than the {\em modified periodogram} for
unevenly-spaced data, so we adopted it to the R light curve as
\cite{fost95} describes.

The observed data $x(t_i)$ can define the data vector
\begin{equation}
    \left|x\right>=[x(t_1),x(t_2),\cdots,x(t_N)].
\end{equation}
First, defining the {\em inner product} of two funcitons $f$ and $g$
as the average value of the product $f^*g$ over the observation
times $\{t_n\}$, we get
\begin{equation}
\left<f|g\right> = (\frac{1}{N})\sum^N_{n=1}f^*(t_n)g(t_n).
\end{equation}
A subspace are spanned by 3 trial functions $\phi_1(t)=1$
(constant), $\phi_2(t)=\sin(\omega t)$, and $\phi_3(t)=\cos(\omega
t)$. These 3 trial functions define a set of trial vectors,
\begin{equation}
    \left|\phi_\alpha\right>=[\phi_\alpha(t_1),\phi_\alpha(t_2),\cdots,\phi_\alpha(t_N)],~\alpha=1,2,3.
\end{equation}
The data vector $\left|x\right>$ can be projected onto the subspace
spanned by the $\left|\phi_\alpha\right>$ results in a model vector
$\left|y\right>$ and a residual vector $\left|\Theta\right>$,
\begin{equation}
    \left|x\right>=\left|y\right>+\left|\Theta\right>,
\end{equation}

The model vector $\left|y\right>$ is defined as
\begin{equation}
\left|y\right>=\sum_{\alpha}c_{\alpha}\left|\phi_{\alpha}\right>.
\end{equation}
The $c_{\alpha}$ can be obtained by taking the inner product of each
trial vector $\phi_\alpha$ with the data vector $x$, and we have
\begin{equation}
\left<\phi_\alpha|x\right>=\sum_{\beta} c_{\beta} \left<\phi_\alpha|\phi_\beta\right>=\sum_\beta S_{\alpha\beta}c_{\beta},
\end{equation}
which defines the $S$ matrix $S_{\alpha\beta}$. Inverting this matrix yields the coefficients,
\begin{equation}
c_\alpha=\sum_\beta S^{-1}_{\alpha\beta}\left<\phi_\beta|x\right>,
\end{equation}
where $s^2$ is the estimated data variance, and it can be replaced
by $\delta^2$. The power level of DCDFT is,
\begin{equation}
P_X(\omega) = \frac{1}{2}N[\left<y|y\right>-\left<1|y\right>^2]/s^2.
\end{equation}

We adopted the {\em false alarm probability}, $F$ \citep{horne86},
to give a quantitative criterion for the detection of a periodic
  signal derived by DCDFT. It was done with the following steps.
First, the power level of the periodogram is normalized by the total variance,
\begin{equation}
P_{N}(\omega) = P_X(\omega)/\delta^2
\end{equation}
The probability that $P_N(\omega_0)$ is of height $z$ or higher is
$Pr[P_{N}(\omega_0)>z]=e^{-z}$. Suppose that $z$ is the highest peak
in a periodogram that samples $N_i$ independent frequencies. The
probability that each independent frequency is smaller than $z$ is
$1-e^{-z}$, so the probability that each frequency is lower than $z$
is $[1-e^{-z}]^{N_i}$. Thus, the false alarm probability (FAP) can
be defined,
\begin{equation}
    F=1-[1-e^{-z}]^{N_i}
\end{equation}
To compute FAP, we need to know $N_i$, which is not too difficult to
obtain by a simple Monte Carlo method.  The FAP tells us the
probability that a peak of height $z$ will occur, assuming that the
data are pure noise. Consequently, the quantity $1-F$ is the
probability that the data contain a signal.

\begin{figure}
    \centering
    \resizebox{\hsize}{!}{\includegraphics*{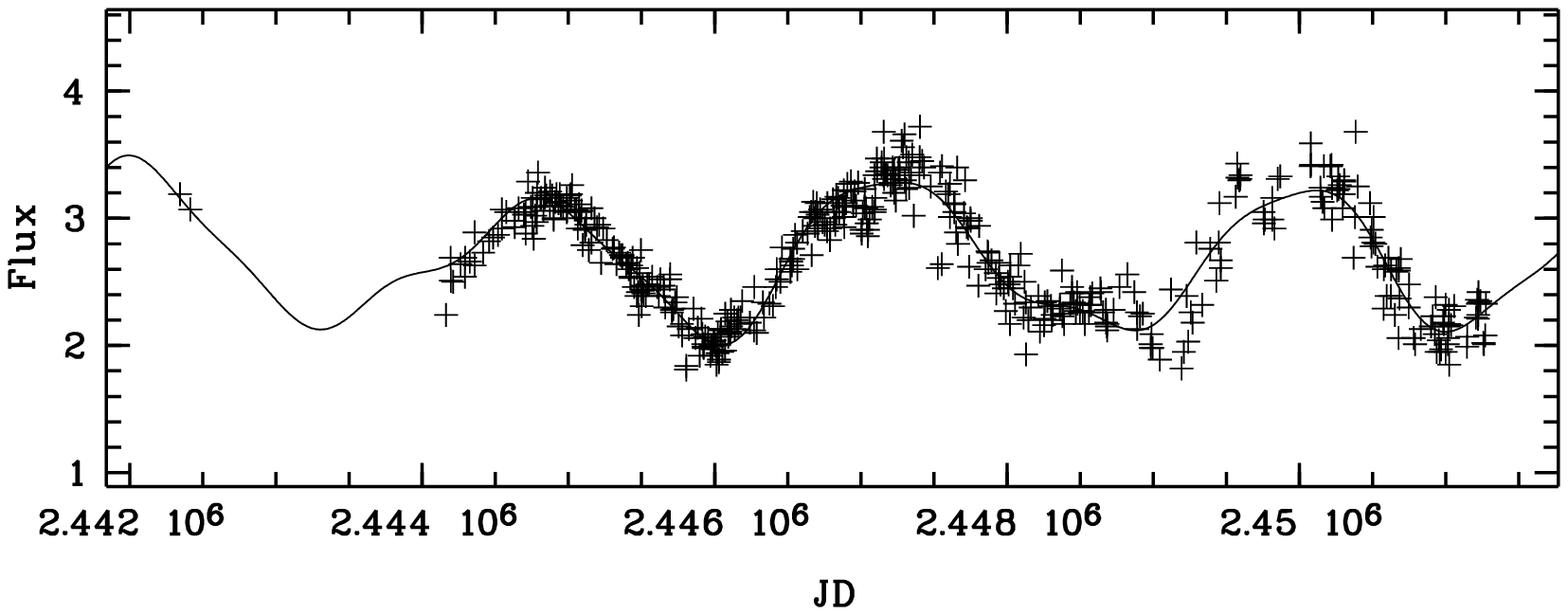}}
    \resizebox{\hsize}{!}{\includegraphics*{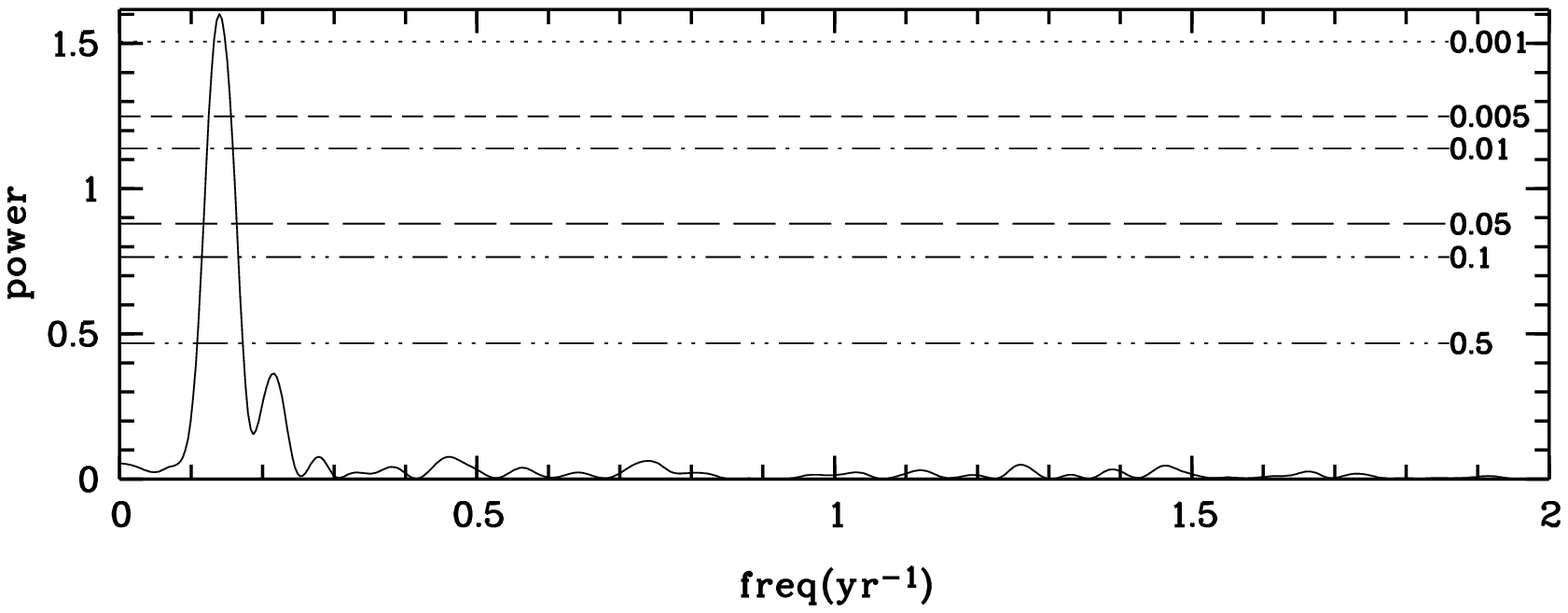}}
    \caption{ The strong sign  of  periods in 0605-085 at
    8GHz. The upper panel is for the light curve at 8GHz, while the lower
    panel is the power spectral analysis result.}
    \label{4775fig1}
\end{figure}

\begin{figure}
    \centering
    \resizebox{\hsize}{!}{\includegraphics*{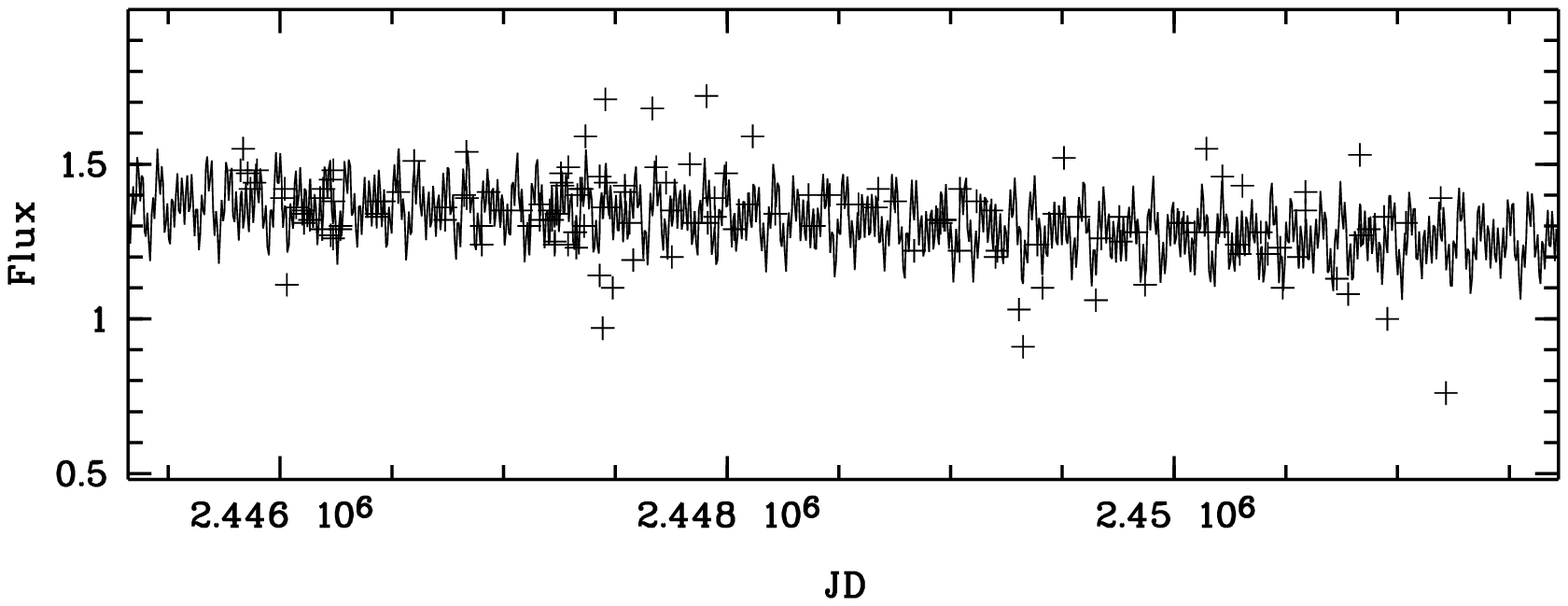}}
    \resizebox{\hsize}{!}{\includegraphics*{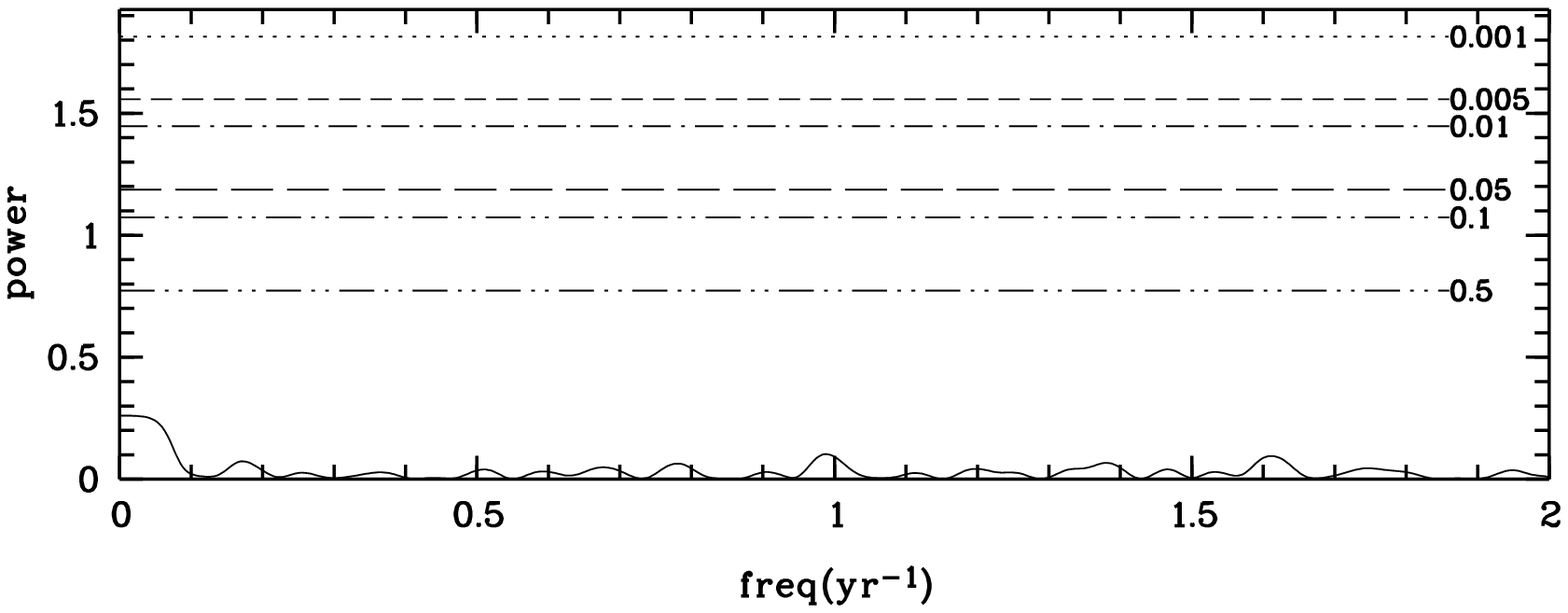}}
    \caption{ The weakest sign  of  periods in 1040+123 at
    8GHz.  The upper panel is for the light curve at 8GHz, while the lower
    panel is the power spectral analysis result.}
    \label{4775fig2}
\end{figure}

For illustration we present the analysis results in Fig.
\ref{4775fig1} for the strong sign of  periods in 0605-085 at 8GHz,
with the theoretical result obtained using two  periods,
  namely a 7.16-year period with an amplitude of
0.582 and a 4.49-year period with an amplitude of 0.168, and Fig.
\ref{4775fig2} for the weakest sign of periods in 1040+123 at 8GHz.

\begin{table*}
\setcounter{table}{0} \caption{Periodicity results and the
variability parameters of radio galaxies. Designation indicates the
name of the source, Freq the frequency in units of GHz, $\delta_N$
the root mean square deviation, the $\delta_N^2$ is the total
variance of data; $A$ -- the amplitude, FAP -- the false alarm
probability of the determined possible period, Tms -- the determined
possible period in units of years, ObT -- the time coverage of the
light curve in units of years, $N$ -- the number of data points, VI
-- VI index, NVA -- NVA index, RMSD -- RMSD index, ID --  source
identification (B for BL Lacertae, Q for the flat spectrum radio
quasar and G for galaxy). The superscript ``F'' means it is not a
physically meaningful period ( $FAP>0.5$ or
$\mathrm{Tms}>\mathrm{ObT}$), and the superscript ``P'' means it is
a possible time scale ($FAP<0.5$ and
$\frac{2}{3}\mathrm{ObT}<\mathrm{Tms}<\mathrm{ObT}$). } \label{tab1}

\begin{center}
\begin{tabular}{@{\hspace{2.5mm}}l@{\hspace{2.5mm}}c@{\hspace{2.5mm}}
r@{\hspace{2.5mm}}r@{\hspace{2.5mm}}r@{\hspace{2.5mm}}r@{\hspace{2.5mm}}
r@{\hspace{2.5mm}}r@{\hspace{2.5mm}}r@{\hspace{2.5mm}}r@{\hspace{2.5mm}}
r@{\hspace{2.5mm}}r@{\hspace{2.5mm}}} \hline\hline Designation &
Freq & $\delta_N$ & A & Fap & Tms & ObT & N & VI & NVA & RMSD & ID
\\ \hline
\object{0003-066}&048& 0.495 & 0.650 & 0.005 &$ 17.9 \pm 3.1 ^F$& 16.7 & 70 & & & & \\
& & 0.495 & 0.648 & 0.006 &$ 9.0 \pm 0.8 $& 16.7 & 70 & & & & \\
& & 0.495 & 0.548 & 0.113 &$ 5.3 \pm 0.5 $& 16.7 & 70 & & & & \\
\object{0003-066}&080& 0.782 & 1.052 & 0.001 &$ 20.6 \pm 2.1 ^F$&
19.7 & 131&
0.93 &0.31 &0.32 &B  \\
& & 0.782 & 0.919 & 0.018 &$ 9.7 \pm 0.9 $& 19.7 & 131 & & & & \\
& & 0.782 & 0.716 & 0.418 &$ 5.7 \pm 0.6 $& 19.7 & 131 & & & & \\
\object{0003-066}&145& 0.739 & 0.997 & 0.002 &$ 29.4 \pm 5.2 ^F$& 17.5 & 102 & & & & \\
\hline
\end{tabular} \end{center} \end{table*}
\clearpage
\begin{table*}
\setcounter{table}{0} \caption{-Continued}
\begin{center}
\begin{tabular}{@{\hspace{2.5mm}}l@{\hspace{2.5mm}}c@{\hspace{2.5mm}}
r@{\hspace{2.5mm}}r@{\hspace{2.5mm}}r@{\hspace{2.5mm}}r@{\hspace{2.5mm}}
r@{\hspace{2.5mm}}r@{\hspace{2.5mm}}r@{\hspace{2.5mm}}r@{\hspace{2.5mm}}
r@{\hspace{2.5mm}}r@{\hspace{2.5mm}}} \hline Designation & Freq &
$\delta_N$ & A & Fap & Tms & ObT & N & VI & NVA & RMSD & ID \\
\hline
\object{0048-097}&048& 0.260 & 0.166 & 0.999 &$ 33.0 \pm 32.3 ^F$& 18.5 & 144 & & & & \\
\object{0048-097}&080& 0.446 & 0.408 & 0.103 &$ 125.2 \pm 109.4 ^F$&
29.1 & 372&
0.92 &0.30 &0.31 &B  \\
\object{0048-097}&145& 0.459 & 0.343 & 0.589 &$ 16.8 \pm 4.4 ^F$&
19.6 & 304\\ \hline
\object{0059+581}&048& 0.339 & 0.312 & 0.988 &$ 0.5 \pm 0.1 ^F$& 4.2 & 27 & & & & \\
\object{0059+581}&080& 0.281 & 0.231 & 1.000 &$ 3.4 \pm 2.4 ^F$& 3.9
& 35&
0.92 &0.10 &0.20 &Q  \\
\object{0059+581}&145& 0.876 & 1.125 & 0.008 &$ 3.4 \pm 0.4 ^P$& 4.8 & 69 & & & & \\
& & 0.876 & 1.101 & 0.013 &$ 1.8 \pm 0.1 $& 4.8 & 69\\
\hline
\end{tabular} \end{center} \end{table*}
\clearpage
\begin{table*}
\setcounter{table}{0} \caption{-Continued}
\begin{center}
\begin{tabular}{@{\hspace{2.5mm}}l@{\hspace{2.5mm}}c@{\hspace{2.5mm}}
r@{\hspace{2.5mm}}r@{\hspace{2.5mm}}r@{\hspace{2.5mm}}r@{\hspace{2.5mm}}
r@{\hspace{2.5mm}}r@{\hspace{2.5mm}}r@{\hspace{2.5mm}}r@{\hspace{2.5mm}}
r@{\hspace{2.5mm}}r@{\hspace{2.5mm}}} \hline Designation & Freq &
$\delta_N$ & A & Fap & Tms & ObT & N & VI & NVA & RMSD & ID \\
\hline
\object{0215+015}&048& 0.253 & 0.216 & 0.871 &$ 15.8 \pm 6.6 ^F$& 18.1 & 85 & & & & \\
\object{0215+015}&080& 0.375 & 0.334 & 0.324 &$ 15.0 \pm 3.3 ^P$&
19.6 & 198&
0.96 &0.24 &0.26 &B  \\
& & 0.375 & 0.341 & 0.270 &$ 3.4 \pm 0.2 $& 19.6 & 198 & & & & \\
\object{0215+015}&145& 0.454 & 0.406 & 0.436 &$ 3.6 \pm 0.2 $& 19.6
& 149\\ \hline
\object{0218+357}&048& 0.080 & 0.066 & 1.000 &$ 4.0 \pm 2.3 ^F$& 6.4 & 27 & & & & \\
\object{0218+357}&080& 0.158 & 0.153 & 0.961 &$ 336.5 \pm 12790.9
^F$& 6.7 & 22&
0.88 &0.11 &0.25 &Q  \\
\object{0218+357}&145& 0.139 & 0.113 & 0.904 &$ 6.7 \pm 4.1 ^F$& 5.0
& 108\\
\hline
\end{tabular} \end{center} \end{table*}
\clearpage
\begin{table*}
\setcounter{table}{0} \caption{-Continued}
\begin{center}
\begin{tabular}{@{\hspace{2.5mm}}l@{\hspace{2.5mm}}c@{\hspace{2.5mm}}
r@{\hspace{2.5mm}}r@{\hspace{2.5mm}}r@{\hspace{2.5mm}}r@{\hspace{2.5mm}}
r@{\hspace{2.5mm}}r@{\hspace{2.5mm}}r@{\hspace{2.5mm}}r@{\hspace{2.5mm}}
r@{\hspace{2.5mm}}r@{\hspace{2.5mm}}} \hline Designation & Freq &
$\delta_N$ & A & Fap & Tms & ObT & N & VI & NVA & RMSD & ID \\
\hline
\object{0336-019}&048& 0.296 & 0.286 & 0.234 &$ 10.0 \pm 1.6 $& 18.8 & 137 & & & & \\
\object{0336-019}&080& 0.400 & 0.406 & 0.021 &$ 10.4 \pm 0.7 $& 24.5
& 491&
0.94 &0.15 &0.16 &Q  \\
\object{0336-019}&145& 0.434 & 0.403 & 0.081 &$ 11.3 \pm 1.0 $& 24.6
& 397\\ \hline
\object{0355+508}&048& 2.655 & 3.698 & 0.000 &$ 925.7 \pm 2033.4 ^F$& 18.5 & 180 & & & & \\
\object{0355+508}&080& 2.981 & 4.076 & 0.000 &$ 33.5 \pm 1.3 ^F$&
29.0 & 666&
* & * & * & \\
\object{0355+508}&145& 3.327 & 4.581 & 0.000 &$ 37.9 \pm 2.2 ^F$&
23.6 & 465\\ \hline
\object{0404+768}&048& 0.097 & 0.084 & 0.999 &$ 0.1 \pm 0.0 ^F$& 8.8 & 34 & & & & \\
\hline
\end{tabular} \end{center} \end{table*}
\clearpage
\begin{table*}
\setcounter{table}{0} \caption{-Continued}
\begin{center}
\begin{tabular}{@{\hspace{2.5mm}}l@{\hspace{2.5mm}}c@{\hspace{2.5mm}}
r@{\hspace{2.5mm}}r@{\hspace{2.5mm}}r@{\hspace{2.5mm}}r@{\hspace{2.5mm}}
r@{\hspace{2.5mm}}r@{\hspace{2.5mm}}r@{\hspace{2.5mm}}r@{\hspace{2.5mm}}
r@{\hspace{2.5mm}}r@{\hspace{2.5mm}}} \hline Designation & Freq &
$\delta_N$ & A & Fap & Tms & ObT & N & VI & NVA & RMSD & ID \\
\hline
\object{0518+165}&048& 0.137 & 0.140 & 0.036 &$ 874.8 \pm 7815.4 ^F$& 17.5 & 312 & & & & \\
\object{0518+165}&080& 0.191 & 0.181 & 0.070 &$ 24.4 \pm 4.7 ^F$&
23.9 & 388&
0.94 &0.25 &0.26 &Q  \\
\object{0518+165}&145& 0.132 & 0.099 & 0.347 &$ 835.1 \pm 10224.4
^F$& 16.7 & 477\\ \hline
\object{0521-365}&048& 0.392 & 0.346 & 0.559 &$ 920.7 \pm 16979.0 ^F$& 18.4 & 127 & & & & \\
\object{0521-365}&080& 0.510 & 0.529 & 0.033 &$ 897.4 \pm 7929.8
^F$& 17.9 & 306&
0.96 &0.08 &0.10 &Q  \\
\object{0521-365}&145& 0.731 & 0.674 & 0.194 &$ 64.0 \pm 53.9 ^F$&
19.5 & 228\\ \hline
\object{0528+134}&048& 1.005 & 1.054 & 0.065 &$ 14.2 \pm 2.5 ^P$& 18.4 & 173 & & & & \\
\hline
\end{tabular} \end{center} \end{table*}
\clearpage
\begin{table*}
\setcounter{table}{0} \caption{-Continued}
\begin{center}
\begin{tabular}{@{\hspace{2.5mm}}l@{\hspace{2.5mm}}c@{\hspace{2.5mm}}
r@{\hspace{2.5mm}}r@{\hspace{2.5mm}}r@{\hspace{2.5mm}}r@{\hspace{2.5mm}}
r@{\hspace{2.5mm}}r@{\hspace{2.5mm}}r@{\hspace{2.5mm}}r@{\hspace{2.5mm}}
r@{\hspace{2.5mm}}r@{\hspace{2.5mm}}} \hline Designation & Freq &
$\delta_N$ & A & Fap & Tms & ObT & N & VI & NVA & RMSD & ID \\
\hline
\object{0710+439}&048& 0.085 & 0.053 & 1.000 &$ 0.1 \pm 0.0 ^F$& 13.9 & 51 & & & & \\
\object{0710+439}&080& 0.133 & 0.088 & 1.000 &$ 0.3 \pm 0.0 ^F$&
21.9 & 58&
0.95 &0.10 &0.17 &Q  \\
\object{0710+439}&145& 0.055 & 0.036 & 1.000 &$ 0.2 \pm 0.0 ^F$&
14.5 & 62\\ \hline
\object{0711+356}&048& 0.127 & 0.142 & 0.051 &$ 711.0 \pm 8723.3 ^F$& 14.2 & 112 & & & & \\
\object{0711+356}&080& 0.134 & 0.087 & 1.000 &$ 805.2 \pm 24638.6
^F$& 16.1 & 111&
0.80 &0.17 &0.22 &Q  \\
\object{0711+356}&145& 0.060 & 0.035 & 1.000 &$ 0.1 \pm 0.0 ^F$&
14.5 & 107\\ \hline
\object{0716+714}&048& 0.215 & 0.191 & 0.283 &$ 15.0 \pm 3.6 ^P$& 17.6 & 219 & & & & \\
\hline
\end{tabular} \end{center} \end{table*}
\clearpage
\begin{table*}
\setcounter{table}{0} \caption{-Continued}
\begin{center}
\begin{tabular}{@{\hspace{2.5mm}}l@{\hspace{2.5mm}}c@{\hspace{2.5mm}}
r@{\hspace{2.5mm}}r@{\hspace{2.5mm}}r@{\hspace{2.5mm}}r@{\hspace{2.5mm}}
r@{\hspace{2.5mm}}r@{\hspace{2.5mm}}r@{\hspace{2.5mm}}r@{\hspace{2.5mm}}
r@{\hspace{2.5mm}}r@{\hspace{2.5mm}}} \hline Designation & Freq &
$\delta_N$ & A & Fap & Tms & ObT & N & VI & NVA & RMSD & ID \\
\hline
\object{0818-128}&048& 0.238 & 0.245 & 0.287 &$ 915.4 \pm 17161.3 ^F$& 18.3 & 70 & & & & \\
\object{0818-128}&080& 0.243 & 0.226 & 0.512 &$ 64.0 \pm 81.0 ^F$&
19.6 & 97&
0.96 &0.26 &0.28 &B  \\
\object{0818-128}&145& 0.268 & 0.250 & 0.491 &$ 8.8 \pm 1.6 $& 18.1
& 99\\ \hline
\object{0829+046}&048& 0.295 & 0.279 & 0.495 &$ 15.0 \pm 4.7 ^P$& 18.6 & 90 & & & & \\
\object{0829+046}&080& 0.445 & 0.486 & 0.033 &$ 15.9 \pm 2.4 ^P$&
21.0 & 187&
0.94 &0.31 &0.32 &B  \\
\object{0829+046}&145& 0.498 & 0.476 & 0.231 &$ 16.8 \pm 4.2 ^P$&
19.6 & 153\\ \hline
\object{0831+557}&048& 0.159 & 0.115 & 1.000 &$ 0.4 \pm 0.0 ^F$& 13.6 & 45 & & & & \\
\hline
\end{tabular} \end{center} \end{table*}
\clearpage
\begin{table*}
\setcounter{table}{0} \caption{-Continued}
\begin{center}
\begin{tabular}{@{\hspace{2.5mm}}l@{\hspace{2.5mm}}c@{\hspace{2.5mm}}
r@{\hspace{2.5mm}}r@{\hspace{2.5mm}}r@{\hspace{2.5mm}}r@{\hspace{2.5mm}}
r@{\hspace{2.5mm}}r@{\hspace{2.5mm}}r@{\hspace{2.5mm}}r@{\hspace{2.5mm}}
r@{\hspace{2.5mm}}r@{\hspace{2.5mm}}} \hline Designation & Freq &
$\delta_N$ & A & Fap & Tms & ObT & N & VI & NVA & RMSD & ID \\
\hline
\object{0917+458}&048& 0.131 & 0.091 & 1.000 &$ 0.3 \pm 0.0 ^F$& 14.2 & 54 & & & & \\
\object{0917+458}&080& 0.130 & 0.095 & 1.000 &$ 0.1 \pm 0.0 ^F$&
15.9 & 60&
0.90 &0.08 &0.16 &Q  \\
\object{0917+458}&145& 0.088 & 0.091 & 0.333 &$ 714.7 \pm 14496.6
^F$& 14.3 & 59\\ \hline
\object{0923+392}&048& 1.870 & 2.616 & 0.000 &$ 24.3 \pm 0.6 ^F$& 21.0 & 476 & & & & \\
\object{0923+392}&080& 2.880 & 3.782 & 0.000 &$ 22.5 \pm 0.7 ^P$&
31.9 &1008&
0.95 &0.32 &0.32 &Q  \\
\object{0923+392}&145& 3.166 & 4.398 & 0.000 &$ 24.4 \pm 0.6 ^P$&
25.0 & 656\\ \hline
\object{0951+699}&048& 0.096 & 0.059 & 1.000 &$ 0.1 \pm 0.0 ^F$& 13.2 & 45 & & & & \\
\hline
\end{tabular} \end{center} \end{table*}
\clearpage
\begin{table*}
\setcounter{table}{0} \caption{-Continued}
\begin{center}
\begin{tabular}{@{\hspace{2.5mm}}l@{\hspace{2.5mm}}c@{\hspace{2.5mm}}
r@{\hspace{2.5mm}}r@{\hspace{2.5mm}}r@{\hspace{2.5mm}}r@{\hspace{2.5mm}}
r@{\hspace{2.5mm}}r@{\hspace{2.5mm}}r@{\hspace{2.5mm}}r@{\hspace{2.5mm}}
r@{\hspace{2.5mm}}r@{\hspace{2.5mm}}} \hline Designation & Freq &
$\delta_N$ & A & Fap & Tms & ObT & N & VI & NVA & RMSD & ID \\
\hline
\object{1101+384}&048& 0.102 & 0.049 & 1.000 &$ 3.7 \pm 0.5 ^F$& 18.4 & 244 & & & & \\
\object{1101+384}&080& 0.127 & 0.068 & 0.958 &$ 8.0 \pm 1.2 ^F$&
21.0 & 416&
0.87 &0.16 &0.19 &B  \\
\object{1101+384}&145& 0.092 & 0.052 & 0.973 &$ 5.4 \pm 0.6 ^F$&
20.0 & 335\\ \hline
\object{1127-145}&048& 0.296 & 0.279 & 0.091 &$ 32.9 \pm 12.9 ^F$& 17.3 & 325 & & & & \\
& & 0.296 & 0.277 & 0.098 &$ 8.7 \pm 0.9 $& 17.3 & 325 & & & & \\
\object{1127-145}&080& 0.515 & 0.591 & 0.001 &$ 1565.8 \pm 6122.6
^F$& 31.3 & 944&
0.96 &0.14 &0.15 &Q  \\
\object{1127-145}&145& 0.289 & 0.204 & 0.377 &$ 3.4 \pm 0.1 $& 24.2
& 577\\
\hline
\end{tabular} \end{center} \end{table*}
\clearpage
\begin{table*}
\setcounter{table}{0} \caption{-Continued}
\begin{center}
\begin{tabular}{@{\hspace{2.5mm}}l@{\hspace{2.5mm}}c@{\hspace{2.5mm}}
r@{\hspace{2.5mm}}r@{\hspace{2.5mm}}r@{\hspace{2.5mm}}r@{\hspace{2.5mm}}
r@{\hspace{2.5mm}}r@{\hspace{2.5mm}}r@{\hspace{2.5mm}}r@{\hspace{2.5mm}}
r@{\hspace{2.5mm}}r@{\hspace{2.5mm}}} \hline Designation & Freq &
$\delta_N$ & A & Fap & Tms & ObT & N & VI & NVA & RMSD & ID \\
\hline
\object{1222+216}&048& 0.201 & 0.220 & 0.914 &$ 2.4 \pm 3.4 ^F$& 1.3 & 12 & & & & \\
\object{1222+216}&080& 0.248 & 0.279 & 0.261 &$ 2.7 \pm 1.6 ^F$& 2.2
& 28&
0.96 &0.11 &0.22 &Q  \\
\object{1222+216}&145& 0.345 & 0.411 & 0.046 &$ 2.6 \pm 0.5 ^P$& 3.5
& 57\\ \hline
\object{1225+206}&048& 0.088 & 0.052 & 1.000 &$ 0.2 \pm 0.0 ^F$& 17.8 & 55 & & & & \\
\object{1225+206}&080& 0.090 & 0.052 & 1.000 &$ 0.2 \pm 0.0 ^F$&
19.6 & 88&
0.55 &0.23 &0.32 &Q  \\
\object{1225+206}&145& 0.034 & 0.020 & 1.000 &$ 64.0 \pm 170.7 ^F$&
19.6 & 78\\ \hline
\object{1226+023}&048& 3.004 & 3.623 & 0.001 &$ 8.8 \pm 0.3 $& 21.0 & 493 & & & & \\
\hline
\end{tabular} \end{center} \end{table*}
\clearpage
\begin{table*}
\setcounter{table}{0} \caption{-Continued}
\begin{center}
\begin{tabular}{@{\hspace{2.5mm}}l@{\hspace{2.5mm}}c@{\hspace{2.5mm}}
r@{\hspace{2.5mm}}r@{\hspace{2.5mm}}r@{\hspace{2.5mm}}r@{\hspace{2.5mm}}
r@{\hspace{2.5mm}}r@{\hspace{2.5mm}}r@{\hspace{2.5mm}}r@{\hspace{2.5mm}}
r@{\hspace{2.5mm}}r@{\hspace{2.5mm}}} \hline Designation & Freq &
$\delta_N$ & A & Fap & Tms & ObT & N & VI & NVA & RMSD & ID \\
\hline
\object{1354-152}&048& 0.122 & 0.158 & 0.556 &$ 0.1 \pm 0.0 ^F$& 5.6 & 6 & & & & \\
\object{1354-152}&080& 0.578 & 0.735 & 0.016 &$ 19.1 \pm 5.3 ^F$&
15.9 & 49&
0.97 &0.32 &0.35 &Q  \\
\object{1354-152}&145& 0.532 & 0.724 & 0.010 &$ 5.7 \pm 0.5 $& 13.8
& 21\\ \hline
\object{1358+624}&048& 0.077 & 0.040 & 1.000 &$ 814.5 \pm 27901.1 ^F$& 16.3 & 153 & & & & \\
\object{1358+624}&080& 0.129 & 0.059 & 1.000 &$ 0.1 \pm 0.0 ^F$&
19.2 & 158&
0.96 &0.09 &0.13 &G  \\
\object{1358+624}&145& 0.054 & 0.023 & 1.000 &$ 0.3 \pm 0.0 ^F$&
16.5 & 136\\ \hline
\object{1400+162}&048& 0.069 & 0.042 & 1.000 &$ 1.0 \pm 0.0 ^F$& 18.7 & 68 & & & & \\
\hline
\end{tabular} \end{center} \end{table*}
\clearpage
\begin{table*}
\setcounter{table}{0} \caption{-Continued}
\begin{center}
\begin{tabular}{@{\hspace{2.5mm}}l@{\hspace{2.5mm}}c@{\hspace{2.5mm}}
r@{\hspace{2.5mm}}r@{\hspace{2.5mm}}r@{\hspace{2.5mm}}r@{\hspace{2.5mm}}
r@{\hspace{2.5mm}}r@{\hspace{2.5mm}}r@{\hspace{2.5mm}}r@{\hspace{2.5mm}}
r@{\hspace{2.5mm}}r@{\hspace{2.5mm}}} \hline Designation & Freq &
$\delta_N$ & A & Fap & Tms & ObT & N & VI & NVA & RMSD & ID \\
\hline
\object{1538+149}&048& 0.252 & 0.284 & 0.050 &$ 29.5 \pm 11.7 ^F$& 18.5 & 103 & & & & \\
\object{1538+149}&080& 0.334 & 0.343 & 0.030 &$ 22.4 \pm 3.9 ^F$&
21.6 & 361&
0.96 &0.22 &0.23 &B  \\
& & 0.334 & 0.322 & 0.064 &$ 10.4 \pm 1.0 $& 21.6 & 361 & & & & \\
\object{1538+149}&145& 0.406 & 0.443 & 0.021 &$ 22.3 \pm 4.4 ^F$& 19.6 & 254 & & & & \\
& & 0.406 & 0.363 & 0.225 &$ 9.4 \pm 1.1 $& 19.6 & 254\\ \hline
\object{1543+005}&048& 0.201 & 0.235 & 0.080 &$ 20.5 \pm 9.2 ^F$& 15.0 & 48 & & & & \\
\object{1543+005}&080& 0.194 & 0.206 & 0.198 &$ 20.6 \pm 7.3 ^F$&
20.4 & 72&
* & * & * & \\
\hline
\end{tabular} \end{center} \end{table*}
\clearpage
\begin{table*}
\setcounter{table}{0} \caption{-Continued}
\begin{center}
\begin{tabular}{@{\hspace{2.5mm}}l@{\hspace{2.5mm}}c@{\hspace{2.5mm}}
r@{\hspace{2.5mm}}r@{\hspace{2.5mm}}r@{\hspace{2.5mm}}r@{\hspace{2.5mm}}
r@{\hspace{2.5mm}}r@{\hspace{2.5mm}}r@{\hspace{2.5mm}}r@{\hspace{2.5mm}}
r@{\hspace{2.5mm}}r@{\hspace{2.5mm}}} \hline Designation & Freq &
$\delta_N$ & A & Fap & Tms & ObT & N & VI & NVA & RMSD & ID \\
\hline
\object{1652+398}&048& 0.150 & 0.158 & 0.052 &$ 933.4 \pm 9878.5 ^F$& 18.7 & 202 & & & & \\
& & 0.150 & 0.155 & 0.065 &$ 11.8 \pm 1.6 $& 18.7 & 202 & & & & \\
\object{1652+398}&080& 0.186 & 0.199 & 0.012 &$ 1064.2 \pm 7297.2
^F$& 21.3 & 440&
0.98 &0.13 &0.15 &B  \\
& & 0.186 & 0.181 & 0.040 &$ 12.3 \pm 1.2 $& 21.3 & 440 & & & & \\
& & 0.186 & 0.146 & 0.288 &$ 7.4 \pm 0.6 $& 21.3 & 440 & & & & \\
\object{1652+398}&145& 0.151 & 0.169 & 0.015 &$ 1001.3 \pm 8145.9 ^F$& 20.0 & 253 & & & & \\
& & 0.151 & 0.158 & 0.038 &$ 11.8 \pm 1.3 $& 20.0 & 253 & & & & \\
\hline
\end{tabular} \end{center} \end{table*}
\clearpage
\begin{table*}
\setcounter{table}{0} \caption{-Continued}
\begin{center}
\begin{tabular}{@{\hspace{2.5mm}}l@{\hspace{2.5mm}}c@{\hspace{2.5mm}}
r@{\hspace{2.5mm}}r@{\hspace{2.5mm}}r@{\hspace{2.5mm}}r@{\hspace{2.5mm}}
r@{\hspace{2.5mm}}r@{\hspace{2.5mm}}r@{\hspace{2.5mm}}r@{\hspace{2.5mm}}
r@{\hspace{2.5mm}}r@{\hspace{2.5mm}}} \hline Designation & Freq &
$\delta_N$ & A & Fap & Tms & ObT & N & VI & NVA & RMSD & ID \\
\hline
\object{1749+701}&048& 0.281 & 0.323 & 0.015 &$ 29.5 \pm 8.0 ^F$& 18.7 & 193 & & & & \\
\object{1749+701}&080& 0.335 & 0.365 & 0.018 &$ 22.3 \pm 4.2 ^F$&
19.2 & 292&
0.67 &0.33 &0.35 &B  \\
& & 0.335 & 0.330 & 0.067 &$ 12.3 \pm 1.6 $& 19.2 & 292 & & & & \\
\object{1749+701}&145& 0.232 & 0.249 & 0.029 &$ 950.8 \pm 8743.0
^F$& 19.0 & 243\\ \hline
\object{1803+784}&048& 0.342 & 0.298 & 0.329 &$ 11.3 \pm 2.0 $& 18.1 & 220 & & & & \\
& & 0.342 & 0.289 & 0.418 &$ 7.8 \pm 1.0 $& 18.1 & 220 & & & & \\
\object{1803+784}&080& 0.409 & 0.236 & 0.997 &$ 1.4 \pm 0.1 ^F$&
18.0 & 233&
0.97 &0.13 &0.15 &B  \\
\hline
\end{tabular} \end{center} \end{table*}
\clearpage
\begin{table*}
\setcounter{table}{0} \caption{-Continued}
\begin{center}
\begin{tabular}{@{\hspace{2.5mm}}l@{\hspace{2.5mm}}c@{\hspace{2.5mm}}
r@{\hspace{2.5mm}}r@{\hspace{2.5mm}}r@{\hspace{2.5mm}}r@{\hspace{2.5mm}}
r@{\hspace{2.5mm}}r@{\hspace{2.5mm}}r@{\hspace{2.5mm}}r@{\hspace{2.5mm}}
r@{\hspace{2.5mm}}r@{\hspace{2.5mm}}} \hline Designation & Freq &
$\delta_N$ & A & Fap & Tms & ObT & N & VI & NVA & RMSD & ID \\
\hline
\object{1939+605}&048& 0.066 & 0.044 & 1.000 &$ 0.3 \pm 0.0 ^F$& 13.9 & 55 & & & & \\
\object{1939+605}&080& 0.142 & 0.101 & 1.000 &$ 13.4 \pm 10.5 ^F$&
13.7 & 51&
0.94 &0.13 &0.21 &G  \\
\object{1939+605}&145& 0.041 & 0.026 & 1.000 &$ 0.1 \pm 0.0 ^F$&
14.3 & 63\\ \hline
\object{1951+498}&048& 0.056 & 0.037 & 1.000 &$ 0.1 \pm 0.0 ^F$& 11.0 & 47 & & & & \\
\object{1951+498}&080& 0.131 & 0.083 & 1.000 &$ 5.5 \pm 2.5 ^F$&
11.1 & 52&
0.71 &0.48 &0.58 &Q  \\
\object{1951+498}&145& 0.044 & 0.028 & 1.000 &$ 0.3 \pm 0.0 ^F$&
11.0 & 45\\ \hline
\object{1954+513}&048& 0.187 & 0.168 & 0.883 &$ 747.0 \pm 19864.3 ^F$& 15.0 & 58 & & & & \\
\hline
\end{tabular} \end{center} \end{table*}
\clearpage
\begin{table*}
\setcounter{table}{0} \caption{-Continued}
\begin{center}
\begin{tabular}{@{\hspace{2.5mm}}l@{\hspace{2.5mm}}c@{\hspace{2.5mm}}
r@{\hspace{2.5mm}}r@{\hspace{2.5mm}}r@{\hspace{2.5mm}}r@{\hspace{2.5mm}}
r@{\hspace{2.5mm}}r@{\hspace{2.5mm}}r@{\hspace{2.5mm}}r@{\hspace{2.5mm}}
r@{\hspace{2.5mm}}r@{\hspace{2.5mm}}} \hline Designation & Freq &
$\delta_N$ & A & Fap & Tms & ObT & N & VI & NVA & RMSD & ID \\
\hline
\object{2134+004}&048& 0.521 & 0.680 & 0.001 &$ 20.6 \pm 1.9 ^F$& 19.5 & 244 & & & & \\
\object{2134+004}&080& 0.946 & 1.195 & 0.000 &$ 86.3 \pm 16.1 ^F$&
31.7 & 603&
0.99 &0.10 &0.11 &Q  \\
\object{2134+004}&145& 0.588 & 0.564 & 0.044 &$ 1218.8 \pm 10105.0
^F$& 24.4 & 474\\ \hline
\object{2136+141}&048& 0.246 & 0.324 & 0.012 &$ 522.6 \pm 5636.5 ^F$& 10.5 & 36 & & & & \\
\object{2136+141}&080& 0.502 & 0.652 & 0.011 &$ 20.7 \pm 4.2 ^P$&
21.0 & 49&
0.89 &0.27 &0.31 &Q  \\
\object{2136+141}&145& 0.394 & 0.492 & 0.022 &$ 9.0 \pm 1.8 ^P$&
11.0 & 50\\ \hline
\object{2145+067}&048& 1.260 & 1.753 & 0.000 &$ 33.0 \pm 2.4 ^F$& 18.2 & 242 & & & & \\
\hline
\end{tabular} \end{center} \end{table*}
\clearpage
\begin{table*}
\setcounter{table}{0} \caption{-Continued}
\begin{center}
\begin{tabular}{@{\hspace{2.5mm}}l@{\hspace{2.5mm}}c@{\hspace{2.5mm}}
r@{\hspace{2.5mm}}r@{\hspace{2.5mm}}r@{\hspace{2.5mm}}r@{\hspace{2.5mm}}
r@{\hspace{2.5mm}}r@{\hspace{2.5mm}}r@{\hspace{2.5mm}}r@{\hspace{2.5mm}}
r@{\hspace{2.5mm}}r@{\hspace{2.5mm}}} \hline Designation & Freq &
$\delta_N$ & A & Fap & Tms & ObT & N & VI & NVA & RMSD & ID \\
\hline
\object{2223-052}&048& 0.764 & 0.784 & 0.025 &$ 17.9 \pm 2.7 ^P$& 19.0 & 406 & & & & \\
& & 0.764 & 0.590 & 0.354 &$ 7.5 \pm 0.8 $& 19.0 & 406 & & & & \\
\object{2223-052}&080& 1.333 & 1.175 & 0.060 &$ 26.9 \pm 3.6 ^P$&
31.8 & 686&
0.96 &0.26 &0.27 &Q  \\
& & 1.333 & 1.143 & 0.077 &$ 7.2 \pm 0.3 $& 31.8 & 686 & & & & \\
\object{2223-052}&145& 1.632 & 1.200 & 0.285 &$ 19.2 \pm 4.2 ^P$& 19.5 & 602 & & & & \\
& & 1.632 & 1.420 & 0.082 &$ 6.5 \pm 0.4 $& 19.5 & 602\\ \hline
\object{2229+391}&048& 0.104 & 0.072 & 1.000 &$ 0.2 \pm 0.0 ^F$& 14.5 & 55 & & & & \\
\hline
\end{tabular} \end{center} \end{table*}
\clearpage
\begin{table*}
\setcounter{table}{0} \caption{-Continued}
\begin{center}
\begin{tabular}{@{\hspace{2.5mm}}l@{\hspace{2.5mm}}c@{\hspace{2.5mm}}
r@{\hspace{2.5mm}}r@{\hspace{2.5mm}}r@{\hspace{2.5mm}}r@{\hspace{2.5mm}}
r@{\hspace{2.5mm}}r@{\hspace{2.5mm}}r@{\hspace{2.5mm}}r@{\hspace{2.5mm}}
r@{\hspace{2.5mm}}r@{\hspace{2.5mm}}} \hline Designation & Freq &
$\delta_N$ & A & Fap & Tms & ObT & N & VI & NVA & RMSD & ID \\
\hline
\object{2352+495}&048& 0.094 & 0.057 & 1.000 &$ 0.4 \pm 0.0 ^F$& 14.5 & 57 & & & & \\
\object{2352+495}&080& 0.140 & 0.101 & 1.000 &$ 62.3 \pm 208.8 ^F$&
13.7 & 58&
0.85 &0.11 &0.19 &Q  \\
\object{2352+495}&145& 0.095 & 0.077 & 0.996 &$ 714.7 \pm 22042.4
^F$& 14.3 & 61\\ \hline
\object{2356+196}&048& 0.135 & 0.156 & 0.900 &$ 0.7 \pm 1.0 ^F$& 0.7 & 3 & & & & \\
\object{2356+196}&080& 0.102 & 0.100 & 0.900 &$ 0.2 \pm 0.0 ^F$&
16.9 & 26&
0.86 &0.13 &0.27 &Q  \\
\object{2356+196}&145& 0.045 & 0.039 & 1.000 &$ 0.1 \pm 0.0 ^F$& 2.0 & 22 & & & & \\
\hline\hline \end{tabular} \end{center} \end{table*}

\section{Variability parameter}

Blazars are variable in the entire electromagnetic wavebands. The
variability violence can be expressed by the variability parameter.
In the optical bands, the variability violence can be reported by
using the variation parameters, such as variability parameter $C$
introduced by \cite{rome99}. In radio bands, the variability
violence is discussed using the variability index ($VI$), the
normalized variability amplitude($NVA$), and the root mean square
dispersion($RMSD$). We described these radio variation parameters
below.

\subsection{Variability index(VI)}

The variability index that measures the peak-to-trough variations in
our flux density measurements can be calculated as introduced by
\cite{aller92,aller03, ciar04}:
\begin{eqnarray} VI =
{\frac{(S_{max}-\sigma_{S_{max}})-(S_{min}+\sigma_{S_{min}})}{(S_{max}-\sigma_{S_{max}})+(S_{min}+\sigma_{S_{min}})}},
\end{eqnarray}
where $S_{max}$ and $S_{min}$ are the peak and the lowest flux
densities, and $\sigma_{\rm{S}_{max}}$ and $\sigma_{\rm{S}_{min}}$
are the associated measurement errors of the fluxes.

\subsection{Normalized variability amplitude (NVA)}

The normalized variability amplitude (NVA) was calculated in the
following way. For each band, the mean $\left<X\right>$ and standard
deviation $\sigma_{\rm{tot}}$ of the flux points and the mean error
level $\sigma_{\rm{err}}$ were determined \citep{edel96}. Since the
NVA is free of the instrumental effect, it is calculated as
\begin{equation}
NVA =
\sqrt{\frac{\sigma_{\rm{tot}}^{2}-\sigma^2_{\rm{err}}}{\left<X\right>^2}}.
\end{equation}

\subsection{Root mean square dispersion (RMSD)}

When a source has been observed at several epochs, whether the
variability is a real one not can be determined by comparing the
distribution of flux at the different epochs with a model in which
the flux of the source is assumed to be non-variable \citep{edel92}.
If $S_{i}$ represents the measured fluxes, then the mean flux
$\left<S\right>$ and the root mean square dispersion as a fraction
of the mean are given by following equations, respectively,
\begin{eqnarray}
    &\left<S\right> &= {\frac{1}{N}}\sum_{i=1}^{n} S_{i} \\
    &\sigma&={\frac{1}{\left<S\right>}}\sqrt{\frac{1}{N-1}\sum_{i=1}^{n} (S_{i}-\left<S\right>)^2} \\
    &\chi^2&={\frac{1}{N}}\sum_{i=1}^{n}({\frac{S_{i}-\left<S\right>}{\sigma_{i}}})^2
\end{eqnarray}
where $\sigma_{i}$ is the uncertainty  in the individual
measurement, and $\chi^2>1$ indicates that the assumption of
non-variable flux is questionable \citep{kemb99}.


\section{Results}

Periodicity results and the variability parameters of radio galaxies
are reported in  Table \ref{tab1}, in which
 Col. 1 represents the name of the source,
 Col. 2 $Freq, $ the frequency  in units of GHz,
 Col. 3 $\delta_N$, root mean square deviation, $\delta_N^2$ is the total variance of data,
 Col. 4 $A$, the amplitude,
 Col. 5 the FAP of the determined  period.
 Col. 6 Tms, the determined period in units of years.
         The superscript ``F'' means it is not a physically meaningful
         period ( $FAP>0.5$ or $\mathrm{Tms}>\mathrm{ObT}$), and the
         superscript ``P'' means it is a possible time scale ($FAP<0.5$
         and $\frac{2}{3}\mathrm{ObT}<\mathrm{Tms}<\mathrm{ObT}$).
 Column. 7 ObT indicates the time coverage of the light curve in units of years,
 Col. 8 the data points ($N$),
 Col. 9 VI,
 Col. 10 NVA,
 Col. 11 RMSD, and
 Col. 12  the source identification (B for BL Lacertae, Q for flat spectrum radio quasar-FSRQ and G for galaxy).

For the whole and the subclass samples, the corresponding averaged
values for the variability parameters are presented in Table
\ref{fan-VI-tab-vp}. The results based on the mutual correlation
between different variability parameters are reported in the Table
\ref{fan-VI-tab-vpc} and plotted in Fig. \ref{4775fig3}.

%
The relationship between the variability parameters and brightness
of the sources were determined by using the averaged 8 GHz flux
density. The results are listed in Table \ref{fan-vi-tab-vb} and the
corresponding results are shown in the Fig. \ref{4775fig4}.
%

\subsection{Periodicity in the radio light curves: Individual source}

The long-term  variability period
 analysis was done at other
wavebands for some sources in the literature. Here we   compare the
radio results with published optical and other EM band results.

\subsubsection{\object{PKS 0219+428} (\object{3C 66A})}

In optical bands, a 65-day  period was reported by \cite{lain99}.
Long-term variability periods of 2.5 years \citep{belo03} and
4.25$\pm$0.28 years \citep{fan02a} were reported.  In the present
paper, no possible period can be found in the period analysis of the
radio light curves.

\subsubsection{\object{AO 0235+164}}

A possible periodicity of $\sim$ 5.7 years was reported in the radio
light curve by \cite{roy00} and \cite{rait01}.  Our result shows
that there is a  period
  of $5.7\pm0.3$ years in 8GHz with
$FAP=0.452$ and $5.8\pm0.3$ years in 14.5GHz with $FAP=0.436$, which
are quite consistent with the earlier result.

\subsubsection{\object{S5 0716+714}}

A periodicity of 5.5$-$6 years was found in the radio emission by
\cite{rait03}.  Our result shows  periods of  $5.7\pm0.5$ years in
4.8GHz with $FAP=0.200$ and $5.4\pm0.3$ years in 14.5GHz with
$FAP=0.117$, which are consistent with the earlier result.

\subsubsection{\object{PKS 0735+178}}

We report  periods of 4.89 and 14.2 years in the optical band (Fan
et al. 1997). The period of 4.89 years was also found by Webb et
al.(1988). For the radio data, a period of $13.5\pm0.9$ years in
14.5GHz with $FAP=0.003$ and    possible periods of
 $12.9\pm0.9$ years in 4.8GHz with $FAP=0.002$ and $15.0\pm1.2$ years in
8GHz with $FAP=0.011$   were found, which are quite consistent with
  the  14.2-year optical period.

\subsubsection{\object{PKS 0754+100}}

The  periods of $3.0\pm0.35$ and 17.85  years were found in our
earlier paper \citep{fan02a}. In the present paper,  periods of
$6.6\pm0.7$ years in 4.8GHz with $FAP=0.222$ and $6.8\pm0.6$ years
in 14.5GHz with $FAP=0.278$, $11.8\pm2.3$ years in 4.8 GHz with
$FAP=0.182$,  $10.4\pm0.8$ years in 8GHz with $FAP=0.014$,
$11.3\pm1.0$ years in 14.5GHz with $FAP=0.015$, and a  possible
 period of  $15.0\pm3.6$ years in 4.8GHz with $FAP=0.167$ were also found. The
$15.0\pm3.6$ year  possible period is consistent with the optical
result, 17.85 years.

\subsubsection{\object{PKS 0851+202} (\object{OJ 287})}

\cite{sill88} reported a 11.65-year  period in the optical light
curve. Periods of $5.53\pm0.15$ and 11.75$\pm$0.5 years were
reported in our earlier paper \citep{fan02a}. But the radio light
curve shows
  periods of
  $8.8\pm1.0$ years in 4.8GHz with
 $FAP=0.445$ and $9.4\pm0.6$ years in 8GHz with $FAP=0.266$, which is
very  different from the previously reported optical periods.

\subsubsection{\object{PKS 1219+285}}

A 14.85$\pm$1.55 year  period was found in the optical band
\citep{fan02a}. The present result shows a  period of $10.4\pm1.4$
years in 8GHz with $FAP=0.293$ and $10.0\pm1.4$ years in 14.5GHz
with $FAP=0.306$.

\subsubsection{\object{PKS 1226+023} (\object{3C 273})}

Periods of 2.0 years and 13.65$\pm$0.2 years were reported in the
optical band. A possible period of 13.5 years was reported in the
X-ray band by \cite{manc02}. The present work gives
  periods of $8.8\pm0.3$ years
 in 4.8GHz with $FAP=0.001$, $8.3\pm0.2$ years in
8GHz with $FAP=0.006$, and $8.2\pm0.2$ years in 14.5GHz with
$FAP=0.001$.

\subsubsection{\object{PKS 1253-055} (\object{3C 279})}

The infrared light curve shows
  a  period
 of 7.1$\pm$0.44 years
\citep{fan99}.  The present work shows
  periods of $5.1\pm0.2$ years with $FAP=0.357$, $7.4\pm0.3$ years with $FAP=0.307$,
$10.1\pm0.7$ years with $FAP=0.426$, and $15.1\pm1.5$ years with
$FAP=0.352$ in 8.0GHz.  The $7.4\pm$0.3 year period is quite
consistent with
 what is
  found in the infrared band.

\subsubsection{\object{PKS 2155-304}}

Based on the optical light curves,  periodicity of 4.6 and 7.0 years
was reported \citep{fan00}. In the present paper, no possible period
was found by power spectral analysis.

\subsubsection{\object{PKS 2200+420} (BL Lacertae)}

The optical periodicity was analyzed and found to be 14.0 years
\citep{fan98}. The present work reports   periods of $3.9\pm0.2$
years in 4.8GHz with $FAP=0.471$, $3.8\pm0.1$ years in 8GHz with
$FAP=0.189$, $3.9\pm0.1$ years in 14.5GHz with $FAP=0.350$,
$7.8\pm0.4$ years in 4.8GHz with $FAP=0.021$, $6.8\pm0.3$ years in
8GHz with $FAP=0.332$, and $7.8\pm0.4$ years in 14.5GHz with
$FAP=0.118$.


\begin{table}
\caption{ Averaged variability parameters} \centering
\begin{tabular}{lccc}
\hline\hline\noalign{\scriptsize}
 $Param.$ & 14.5GHz & 8GHz &  4.8GHz \\
(1)     & (2)  & (3)  & (4)  \\
\noalign{\smallskip} \hline
VI      &   0.92   $\pm$   0.08   &   0.90   $\pm$   0.09   &   0.90   $\pm$   0.10   \\
VI-BL   &   0.90   $\pm$   0.10   &   0.88   $\pm$   0.12   &   0.84   $\pm$   0.14   \\
VI-FSRQ &   0.94   $\pm$   0.05   &   0.92   $\pm$   0.07   &   0.93   $\pm$   0.06   \\
VI-G    &   0.89   $\pm$   0.09   &   0.90   $\pm$   0.06   & 0.93
$\pm$ 0.05
\\\hline
NVA     &   0.23   $\pm$   0.13   &   0.22   $\pm$   0.12   &   0.16   $\pm$   0.12   \\
NVA-BL  &   0.30   $\pm$   0.12   &   0.28   $\pm$   0.11   &   0.24   $\pm$   0.11   \\
NVA-FSRQ&   0.22   $\pm$   0.11   &   0.21   $\pm$   0.11   &   0.15   $\pm$   0.11   \\
NVA-G   &   0.10   $\pm$   0.04   &   0.11   $\pm$   0.05   &   0.04
$\pm$ 0.03
\\\hline
RMSD    &   0.27   $\pm$   0.12   &   0.27   $\pm$   0.11   &   0.23   $\pm$   0.12   \\
RMSD-BL &   0.33   $\pm$   0.11   &   0.32   $\pm$   0.11   &   0.29   $\pm$   0.11   \\
RMSD-FSRQ&  0.25   $\pm$   0.10   &   0.24   $\pm$   0.10   &   0.20   $\pm$   0.10   \\
RMSD-G  &   0.17   $\pm$   0.03   &   0.20   $\pm$   0.05   &   0.15   $\pm$ 0.03     \\
\hline\hline
\end{tabular}\\
\label{fan-VI-tab-vp}
\end{table}

\begin{table}
\caption{  Correlation between variability parameters, $Y=aX+b$}
\centering
\begin{tabular}{lccc}
\hline\hline\noalign{\scriptsize}
 $X-Y$ & $a \pm \Delta a$ & b $\pm \Delta b$ &  $r$ \\
(1)     & (2)  & (3)  & (4)  \\
\noalign{\smallskip} \hline
V.I.--NVA &  $0.04\pm0.06$ &  0.89$\pm$ 0.02   &  0.05   \\
RMSD-NVA  &   $0.84\pm0.03$ &  1.08$\pm$ 0.01   &  0.893   \\
\hline\hline
\end{tabular}\\
\label{fan-VI-tab-vpc}
\end{table}

\begin{figure}
    \centering
    \resizebox{\hsize}{!}{\includegraphics*{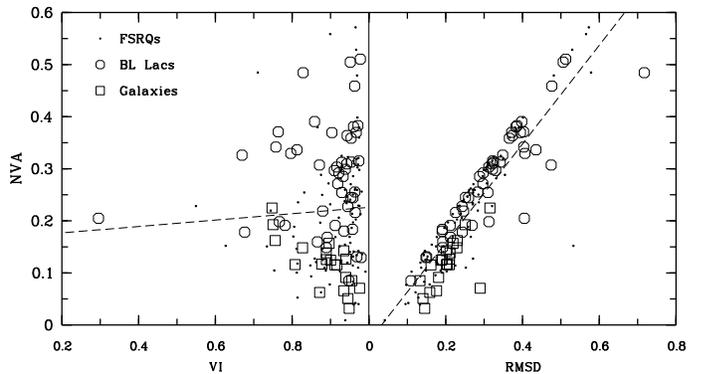}}
    \caption{ Correlation between variability parameters. Left panel is for
    NVA and V.I., while the right panel is for NVA and RMSD.}
    \label{4775fig3}
\end{figure}

\begin{table}
\caption{  Correlation between variability parameter and flux
density, $\mathrm{Vari}=a\log F_{\mathrm{8GHz}} + b$} \centering
\begin{tabular}{lccc}
\hline\hline\noalign{\scriptsize}
 $Param.$ & $a \pm \Delta a$ & $b \pm \Delta b$ &  $r$ \\
(1)     & (2)  & (3)  & (4)  \\
\noalign{\smallskip} \hline
log $F_{\mathrm{8GHz}}-RMSD$ &  $-0.04\pm0.02$ &  2.27$\pm$ 0.01   &  -0.17   \\
log $F_{\mathrm{8GHz}}-NVA$   &   $0.01\pm0.02$ &  1.41$\pm$ 0.01   &  0.05   \\
log $F_{\mathrm{8GHz}}-VI$    &   $0.132\pm0.01$ &  0.88$\pm5.7\times10^{-3}$   &  0.65   \\
\hline\hline
\end{tabular}\\
\label{fan-vi-tab-vb}
\end{table}

\begin{figure}
    \centering
    \resizebox{\hsize}{!}{\includegraphics*{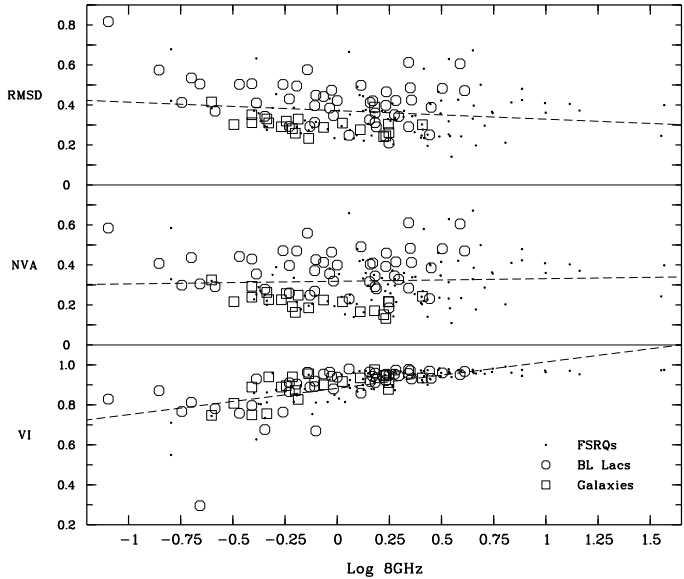}}
    \caption{ Correlation between the variability parameter and radio
    brightness (averaged flux density). The upper panel indicates  the
    relation between RMSD and the 8GHz flux density, the middle panel
    indicates the relation between NVA and the 8GHz flux density, and the
    lower panel indicates  the relation between V.I. and the 8GHz flux
    density. The lines show the best fitting results mentioned in the
    text.}
    \label{4775fig4}
\end{figure}

\section{Discussions and conclusions}

Blazars are variable over all electromagnetic wavelengths. Optical
photometry is available for some blazars for about a century
\citep{fan05a}. The radio monitoring program started much too late,
only about 40 years ago. But the radio monitoring coverage is also
long enough for the periodicity analysis and the variability
property investigations.

There are 168 radio sources in  the UMRAO data base, and the
observation time coverage of the radio light curves is from 0.1 year
for \object{1217+023} to 33.8 years for \object{1226+023}
(\object{3C 273}).  When the power spectral periodicity analysis
method was adopted to the  4.8GHz, 8.0GHZ, and 14.5GHz light curves
for the 168 sources,  203 astrophysically
  meaningful  periods  ( $FAP <
0.50$ and $\mathrm{Tms}<\frac{2}{3}\mathrm{ObT}$) were obtained for
66 sources (see Table \ref{tab1}). The periods are different from
one source to another, which is from 2.2 years for \object{0454-234}
at 4.8 GHz to 20.8$\pm$1.2 years for \object{1641+399} at 8.0 GHz.
There is no clear  possible period sign found for the other 102
sources for which either the FAP is greater than 0.5 for the period
or the period is 2/3 times longer than the observation time coverage
(see Table \ref{tab1}, the superscript ``F'' means the period is not
a physically meaningful one ($FAP>0.5$ or
$\mathrm{Tms}>\mathrm{ObT}$), and the superscript ``P'' means it is
a possible time scale ($FAP<0.5$ and
$2/3\mathrm{ObT}<\mathrm{Tms}<\mathrm{ObT}$)). Here, we take the
periods with $FAP>0.5$ to have  no physical meaning. In addition, if
the determined  period is $2/3$ times longer than the observation
time coverage, we did not take it as a physically meaningful period
either.  It can be mentioned that from data that is pure noise, any
method of period estimation will yield some false probabilities in
the period range that is roughly equal to the length of the data
samples, or somewhat smaller.

If we consider galaxie, FSRQs and BLs separately, we find that the
physically significant  periodicity at 8GHz are in the range of
 2.2 to 20.8 years for FSRQs (55 physically meaningful  periods
for 34 objects) and from 2.5 to 18.0 years for BLs (27 physically
meaningful periods for 17 objects). However, there is no physically
significant periodicity found for galaxies.  The average value of
the  periodicity is $ 8.9 \pm 4.0$ years for FSRQs and $ 8.1 \pm
3.4$ years for BLs.  In Table \ref{fan-VI-tab-vp} and Fig
\ref{4775fig4}, we can see that RMSD and NVA for galaxies are lower
than those for FSRQs and BLs. It is interesting that the sources
with physically meaningful periods have higher RMSD and NVA.

\begin{figure}
    \centering
    \resizebox{\hsize}{!}{\includegraphics*{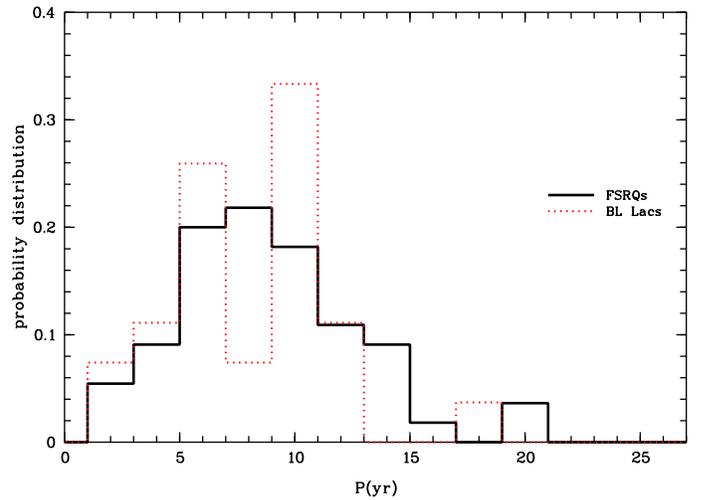}}
    \caption{  Histogram of the periodicity, P (in units of
    years) at 8GHz for BL Lacertae objects and FSRQs. The dotted line stands
    for BLs and the filled line for FSRQs.} \label{4775Fig5}
\end{figure}

\begin{figure}
    \centering
        \resizebox{\hsize}{!}{\includegraphics*{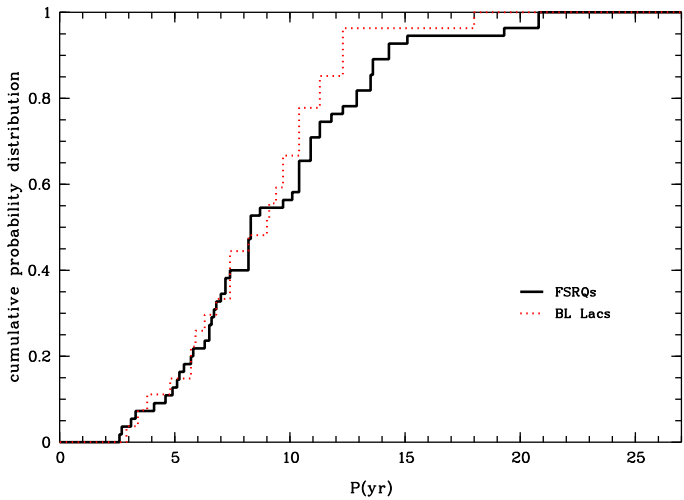}}
    \caption{ Accumulative results for the periodicity (in units of
    years) at 8GHz for BL Lacertae objects and FSRQs. The dotted line stands for
    BLs and filled line for FSRQs.The Kolmogorov-Smirnov test indicates
    that the probability for the possible periodicity distributions of BL
    Lac objects and FSRQs coming from the same parent distribution is 68.2\%. }
    \label{4775fig6}
\end{figure}

Our results are also consistent with the results obtained by
\cite{ciar04}, who have analyzed the  periodicity from the
high-frequency radio data.  The physically meaningful  period
histogram at 8GHz for the subclass samples of BLs and FSRQs are
shown in Fig. \ref{4775Fig5}. There is no clear difference in the
possible periodicity distribution as shown in Fig. \ref{4775fig6},
in which the Kolmogorov-Smirnov (K-S) test indicates that the
probability for the possible periodicity distributions of BL Lac
objects and FSRQs coming from the same parent distribution is
greater than 68.2\%. If the possible periodicity is associated with
the central structure, namely associated with the central black-hole
mass, then the similar possible periodicity distribution for FSRQs
and BLs should suggest that their central black-hole masses   show a
similar distribution. In fact,  no clear difference was found in the
central black-hole masses of BLs and FSRQs \citep{fan05b}.

From the sources listed in Sect. 4.1, we can see that  the periods
found in the optical bands were not always consistent with those
found in the radio bands.
 Some sources show similar radio and
optical variability  periods: 0735+178 show a significant radio
period of $13.5\pm0.9$ years and two possible radio periods of
12.9$\pm$0.9 and 15.0$\pm$1.2, and an optical  period of 14.2 years,
0754+100 shows a  possible radio period of $15.0\pm3.6$ years and an
optical  period of 17.85 years, 1253-055 shows radio  periods of 7.1
to 7.4 years and an infrared  period of $7.1\pm0.44$ years for
instance. Meanwhile some others show different possible periods(OJ
287, 1226+023(3C 273), 2200+420 for instance.
  This difference is perhaps
from the fact that the light curves used for the possible
periodicity analysis were not long enough in some sources, or the
variation in the radio bands and optical bands were caused by
different mechanisms as noticed in the case of OJ 287, and the
observed optical outbursts were not correlated with those observed
in the radio band \citep{taka98}.  For some cases, it is possible
that the lack of agreement between optical and radio possible
periods is due to the fact that one or the other is spurious, just
due to noise.

For AGNs, the variability mechanism is not yet well understood. Some
models have been proposed to explain the optical long-term possible
periodic variations: the binary black-hole model, the thermal
instability model, and the perturbation model \citep{fan05a}. The
promising models are the binary black-hole model and the
perturbation model.  The helical jet related with the binary black
holes have been used to explain the optical variability behavior for
the objects (3C 345, OJ 287, BL Lacertae, and PKS 0735+178). It has
been claimed that the possible periodicity in the historical light
curves also show helical trajectories in their VLBI radio components
\citep{vill99}.  In this sense, one would expect  similar possible
periodicity behavior in optical and radio bands, which has been
confirmed for many sources in our analysis. In the radio bands, the
variability is explained by various mechanisms \citep{ciar04} such
as shocks in jets, changes in the direction of forward beaming, and
precession in a binary black-hole system
\citep{mars85,aller85,came92,bege80, rm00, rm03}.

The variability parameters viz. variability index VI, normalized
variability amplitude NVA, and  RMSD are listed in the Table
\ref{fan-VI-tab-vp}, in which $W$ stands for the whole sample. These
parameters show that the sources are variable, and NVA and RMSD are
correlated. On the other hand, no correlation was found in the VI
parameter with NVA or RMSD (see Table \ref{fan-VI-tab-vpc}, Fig.
\ref{4775fig3}). This suggests that NVA and RMSD are more reliable
for the variability indication than the VI. Therefore, we suggest
using NVA and RMSD for indicating variability violence, if possible.
It is also easily found that the variability parameter at higher
frequency is greater than those in the lower frequency.

However, for the correlation between the variability parameter and
the flux density, we found that the relationship between the source
brightness and NVA and/or RMSD is not as close as the one between
the source flux density and VI (see Fig. \ref{4775fig4} and Table
\ref{fan-vi-tab-vb}). We think that the correlation between the
source brightness and VI is an apparent result, since the VI and the
brightness (namely the averaged flux density) are more associated
with the maximum flux density when the difference between the
maximum and the minimum is big enough, which will result in an
apparent correlation. Therefore, we do not think that there is a
correlation between the brightness and the variability in the radio
bands.

When we considered BLs and FSRQs separately, we found that the NVA
and RMSD of BLs are larger than those of FSRQs,  and the NVA and
RMSD of BLs and FSRQs are  larger than those of galaxies. This
finding  suggests that BLs are more variable than FSRQs in the radio
bands. In addition, there is a tendency for the variability
parameter to increase with the frequency for the whole sample and
the individual BLs and FSRQ sub-samples. This tendency is also found
in the optical bands.

In the present paper, the power spectral (Fourier) periodicity
analysis method was adopted to  a large sample of radio sources
given in UMRAO.  The results show that the possible periodicity is
present in the  range of 2.2 to 20.8 years for 66 radio sources. BLs
are more variable than FSRQs and the variability parameters depend
on the frequency.

\begin{acknowledgements}

The authors thank the referee for the constructive comments and
suggestions. This work is partially supported by the National 973
project (NKBRSF G19990754), the National Science Fund for
Distinguished Young Scholars (10125313), the National Natural
Science Foundation of China (10573005, 10633010), and the Fund for
Top Scholars of Guangdong Province (Q02114). We also thank the
financial support from the Guangzhou Education Bureau and Guangzhou
Science and Technology Bureau. ACG's efforts are partially supported
by the Department of Atomic Energy, Govt. of India. This research
made use of data from the University of Michigan Radio Astronomy
Observatory, which is supported by the University of Michigan and
the National Science Foundation.

\end{acknowledgements}

\end{document}